\documentclass[preprint,prd,aps,showpacs,showkeys,nofootinbib]{revtex4}
\usepackage{graphicx}
\usepackage{dcolumn}
\usepackage{bm}
\usepackage{color}
\usepackage{amsmath}
\usepackage{amssymb}
\begin{document}
\title{The study of lepton EDMs in $U(1)_X$SSM}

\author{Lu-Hao Su$^{1,2}$\footnote{suluhao0606@163.com}, Dan He$^{1,2}$\footnote{hedandeya@163.com}, Xing-Xing Dong$^{1,2}$\footnote{dxx$\_$0304@163.com}, Tai-Fu Feng$^{1,2,3}$\footnote{fengtf@hbu.edu.cn}, Shu-Min Zhao$^{1,2}$\footnote{zhaosm@hbu.edu.cn}}

\affiliation{$^1$ Department of Physics, Hebei University, Baoding 071002, China}
\affiliation{$^2$ Key Laboratory of High-precision Computation and Application of Quantum Field Theory of Hebei Province, Baoding 071002, China}
\affiliation{$^3$ Department of Physics, Chongqing University, Chongqing 401331, China}
\begin{abstract}
  The minimal supersymmetric extension of the standard model (MSSM) is extended to the $U(1)_X$SSM, whose local gauge group is $SU(3)_C \times SU(2)_L \times U(1)_Y \times U(1)_X$. To obtain the $U(1)_X$SSM, we add  the new superfields to the MSSM, namely:
  three Higgs singlets $\hat{\eta},~\hat{\bar{\eta}},~\hat{S}$ and right-handed neutrinos $\hat{\nu}_i$. The CP violating effects are considered to study the lepton electric dipole moment(EDM) in $U(1)_X$SSM. The CP violating phases in $U(1)_X$SSM are more than those in the standard model(SM). In this model, some new parameters $(\theta_S, \theta_{BB^{\prime}}, \theta_{BL})$ as CP violating phases are considered, so there are new contributions to lepton EDMs. It is conducive to exploring the source of CP violation and probing new physical beyond SM.
\end{abstract}


\maketitle

\section{Introduction}
In 1964, Cronin and Fitch discovered the charge conjugate and parity (CP) violating by the decays of the $K$ meson \cite{1964}. The study of the lepton electric dipole moments (EDMs) becomes the physical quantities for probing sources of CP violation\cite{2}.
Therefore, it is of special significance to research the EDMs of lepton. At present, the upper bound of electron EDM is $|d^{exp}_e|$ $<$ $1.1 \times 10^{-29}$ e.cm at the $90 \%$ confidence level\cite{de,de1,de2},  the muon EDM is $|d^{exp}_{\mu}|$ $<$ $1.8 \times 10^{-19}$ e.cm at the $95 \%$ confidence level and the tau EDM is $|d^{exp}_{\tau}|$ $<$ $1.1 \times 10^{-17}$ e.cm at the $95 \%$ confidence level \cite{pdg}. The sources and mechanism of CP violation have not been well explained. Scientists are trying to find CP violation terms in new physics beyond SM in order to better explain CP violation mechanism\cite{NPdl1,NPdl2,NPdl3,NPdl4}. There are several CP violating phases, which can give large contributions to the EDMs of lepton in the minimal supersymmetric extension of standard model (MSSM)\cite{mssm,mssm1,mssm2,Z2015}.

Due to the deficiency of MSSM which can not explain neutrino mass and not solve $\mu$ problem, U(1) extension of MSSM is carried out. There are two U(1) groups in $U(1)_X$SSM: $U(1)_Y$ and $U(1)_X$, and the $U(1)_X$SSM is explored that we use SARAH software packages \cite{extend1, extend2, extend3}. By adding some new superfields to MSSM, the $U(1)_X$SSM not only obtains additional Higgs, neutrino and gauge fields, but corresponding
superpartners that extend the neutralino and sfermion sectors. The mass $m_{h_0}$ of the lightest CP-even Higgs \cite{LCTHiggs1,LCTHiggs2} in $U(1)_X$SSM is larger than the corresponding mass in MSSM at tree order. Therefore, In $U(1)_X$SSM, the loop diagram correction of $m_{h_0}$ does not need to be very large.

It is an effective way to explore new physics beyond the standard model(SM) that research the MDMs \cite{mdms,mdms1} and EDMs \cite{edms1, edms2, edms3, edms4, edms6, edms7, edms8} of lepton. The one-loop correction and the two-loop correction to EDM of leptons are well researched in MSSM. $d_e$ in the SM is studied independently of the model\cite{mi,mi1}. The authors consider the right-handed neutrinos, the neutrino see-saw mechanism and the structure of minimal flavor violation. The results show that when the neutrinos are Majorana particles, the value of $d_e$ will reach the upper limit of the experiment.

In the following, we introduce the specific form of $U(1)_X$SSM and its superfields in section 2. In section 3, we show that the one-loop and two-loop corrections to the lepton EDMs. The main content of section 4 is the numerical analysis for the dependence of lepton EDMs on the $U(1)_X$SSM parameters. We have a special summary and discussion in section 5. The appendix is used for the some mass matrics.

\section{the $U(1)_X$SSM}
The $U(1)_X$SSM has been expanded on the basis of the MSSM. The $U(1)_X$SSM superfields include
three Higgs singlets $\hat{\eta},~\hat{\bar{\eta}},~\hat{S}$ and right-handed neutrinos $\hat{\nu}_i$. By the see-saw mechanism, the light neutrinos can get tiny masses at the tree level. For details of the mass matrix of particles, please see the Ref.\cite{pro}.

The superpotential of $U(1)_X$SSM is
\begin{eqnarray}
&&W=l_W\hat{S}+\mu\hat{H}_u\hat{H}_d+M_S\hat{S}\hat{S}-Y_d\hat{d}\hat{q}\hat{H}_d-Y_e\hat{e}\hat{l}\hat{H}_d+\lambda_H\hat{S}\hat{H}_u\hat{H}_d
\nonumber\\&&+\lambda_C\hat{S}\hat{\eta}\hat{\bar{\eta}}+\frac{\kappa}{3}\hat{S}\hat{S}\hat{S}+Y_u\hat{u}\hat{q}\hat{H}_u+Y_X\hat{\nu}\hat{\bar{\eta}}\hat{\nu}
+Y_\nu\hat{\nu}\hat{l}\hat{H}_u\;.
\end{eqnarray}

These two Higgs doublets and three Higgs singlets are shown below in concrete form,
\begin{eqnarray}
&&H_{u}=\left(\begin{array}{c}H_{u}^+\\{1\over\sqrt{2}}\Big(v_{u}+H_{u}^0+iP_{u}^0\Big)\end{array}\right)\;,
~~~~~~
H_{d}=\left(\begin{array}{c}{1\over\sqrt{2}}\Big(v_{d}+H_{d}^0+iP_{d}^0\Big)\\H_{d}^-\end{array}\right)\;,
\nonumber\\
&&\eta={1\over\sqrt{2}}\Big(v_{\eta}+\phi_{\eta}^0+iP_{\eta}^0\Big)\;,~~~~~~~~~~~~~~~
\bar{\eta}={1\over\sqrt{2}}\Big(v_{\bar{\eta}}+\phi_{\bar{\eta}}^0+iP_{\bar{\eta}}^0\Big)\;,\nonumber\\&&
\hspace{4.0cm}S={1\over\sqrt{2}}\Big(v_{S}+\phi_{S}^0+iP_{S}^0\Big)\;.
\end{eqnarray}
$v_u,~v_d,~v_\eta$,~ $v_{\bar\eta}$ and $v_S$ are the VEVs of the Higgs superfields $H_u$, $H_d$, $\eta$, $\bar{\eta}$ and $S$ respectively.

Here, we set $\tan\beta=v_u/v_d$ and $\tan\beta_\eta=v_{\bar{\eta}}/v_{\eta}$. The specific form of
$\tilde{\nu}_L$ and $\tilde{\nu}_R$ are
\begin{eqnarray}
\tilde{\nu}_L=\frac{1}{\sqrt{2}}\phi_l+\frac{i}{\sqrt{2}}\sigma_l\;,~~~~~~~~~~\tilde{\nu}_R=\frac{1}{\sqrt{2}}\phi_R+\frac{i}{\sqrt{2}}\sigma_R\;.
\end{eqnarray}

The specific form of soft SUSY breaking terms are shown below
\begin{eqnarray}
&&\mathcal{L}_{soft}=\mathcal{L}_{soft}^{MSSM}-B_SS^2-L_SS-\frac{T_\kappa}{3}S^3-T_{\lambda_C}S\eta\bar{\eta}
+\epsilon_{ij}T_{\lambda_H}SH_d^iH_u^j\nonumber\\&&
-T_X^{IJ}\bar{\eta}\tilde{\nu}_R^{*I}\tilde{\nu}_R^{*J}
+\epsilon_{ij}T^{IJ}_{\nu}H_u^i\tilde{\nu}_R^{I*}\tilde{l}_j^J
-m_{\eta}^2|\eta|^2-m_{\bar{\eta}}^2|\bar{\eta}|^2\nonumber\\&&
-m_S^2S^2-(m_{\tilde{\nu}_R}^2)^{IJ}\tilde{\nu}_R^{I*}\tilde{\nu}_R^{J}
-\frac{1}{2}\Big(M_S\lambda^2_{\tilde{X}}+2M_{BB^\prime}\lambda_{\tilde{B}}\lambda_{\tilde{X}}\Big)+h.c.\;.
\end{eqnarray}

The particle content and charge assignments for $U(1)_X$SSM are shown in the Table~\ref{charge}.
Compared to the SM, the anomalies of $U(1)_X$SSM are more complex \cite{text}.
This model has been proven anomaly free \cite{pro} at last.
The two Abelian groups $U(1)_Y$ and $U(1)_X$ in $U(1)_X$SSM can create a new effect,
the gauge kinetic mixing. This effect can also be induced by RGEs even with zero value at $M_{GUT}$.

\begin{table}[h]
  \centering
  \begin{tabular}{|c|c|c|c|c|}
  \hline
  Superfields & $SU(3)_C$ & $SU(2)_L$ & $U(1)_Y$ & $U(1)_X$ \\
  \hline
  $\hat{Q}_i$ & 3 & 2 & 1/6 & 0 \\
  \hline
  $\hat{u}^c_i$ & $\bar{3}$ & 1 & -2/3 & -$1/2$ \\
  \hline
  $\hat{d}^c_i$ & $\bar{3}$ & 1 & 1/3 & $1/2$  \\
  \hline
  $\hat{L}_i$ & 1 & 2 & -1/2 & 0  \\
  \hline
  $\hat{e}^c_i$ & 1 & 1 & 1 & $1/2$  \\
  \hline
  $\hat{\nu}_i$ & 1 & 1 & 0 & -$1/2$ \\
  \hline
  $\hat{H}_u$ & 1 & 2 & 1/2 & 1/2\\
  \hline
  $\hat{H}_d$ & 1 & 2 & -1/2 & -1/2 \\
  \hline
  $\hat{\eta}$ & 1 & 1 & 0 & -1 \\
  \hline
  $\hat{\bar{\eta}}$ & 1 & 1 & 0 & 1\\
  \hline
  $\hat{S}$ & 1 & 1 & 0 & 0 \\
  \hline
  \end{tabular}
  \caption{The superfields in $U(1)_X$SSM.}
  \label{charge}
\end{table}

The general form of the covariant derivative of this model is \cite{model1, model2, model3}
\begin{eqnarray}
&&D_\mu=\partial_\mu-i\left(\begin{array}{cc}Y,&X\end{array}\right)
\left(\begin{array}{cc}g_{Y},&g{'}_{{YX}}\\g{'}_{{XY}},&g{'}_{{X}}\end{array}\right)
\left(\begin{array}{c}A_{\mu}^{\prime Y} \\ A_{\mu}^{\prime X}\end{array}\right)\;.
\label{gauge1}
\end{eqnarray}

The $A_{\mu}^{\prime Y}$ and $A^{\prime X}_\mu$ represent the gauge fields of $U(1)_Y$ and $U(1)_X$. Since these two Abelian gauge groups are unbroken, we make a basis exchange.
Using the orthogonal matrix $R$ \cite{model1, model3}, the resulting formula is shown below
\begin{eqnarray}
&&\left(\begin{array}{cc}g_{Y},&g{'}_{{YX}}\\g{'}_{{XY}},&g{'}_{{X}}\end{array}\right)
R^T=\left(\begin{array}{cc}g_{1},&g_{{YX}}\\0,&g_{{X}}\end{array}\right)\;.
\label{gauge2}
\end{eqnarray}

We deduce $\sin^2\theta_{W}^\prime$ $=$
\begin{eqnarray}
\frac{1}{2}-\frac{((g_{YX}+g_X)^2-g_{1}^2-g_{2}^2)v^2+
4g_{X}^2\xi^2}{2\sqrt{((g_{YX}+g_X)^2+g_{1}^2+g_{2}^2)^2v^4+8g_{X}^2((g_{YX}+g_X)^2-g_{1}^2-g_{2}^2)v^2\xi^2+16g_{X}^4\xi^4}}\;.
\end{eqnarray}
with $\xi=\sqrt{v_\eta^2+v_{\bar{\eta}}^2}$.
The new mixing angle $\theta_{W}^\prime$ can be found in the couplings of $Z$ and $Z^{\prime}$.

Next, we describe some of the couplings needed.

The couplings of $\tilde{\nu}^R_k-\bar{e}_i-\chi_j^-$ and $\tilde{\nu}^I_k-\bar{e}_i-\chi_j^-$ are
\begin{eqnarray}
&&\mathcal{L}_{\tilde{\nu}^R\bar{e}\chi^-}=\bar{e}_i\Big\{\frac{i}{\sqrt{2}}U^*_{j2}Z^{R*}_{ki}Y_e^iP_L-\frac{i}{\sqrt{2}}g_2V_{j1}Z^{R*}_{ki}P_R\Big\}\chi_j^-\tilde{\nu}^R_k\;,
\\&&\mathcal{L}_{\tilde{\nu}^I\bar{e}\chi^-}=\bar{e}_i\Big\{\frac{-1}{\sqrt{2}}U^*_{j2}Z^{I*}_{ki}Y_e^iP_L+\frac{1}{\sqrt{2}}g_2V_{j1}Z^{I*}_{ki}P_R\Big\}\chi_j^-\tilde{\nu}^I_k\;.
\end{eqnarray}
With $P_L=\frac{1-\gamma_5}{2}$ and $P_R=\frac{1+\gamma_5}{2}$. $Z^{R}$ and $Z^{I}$ are rotation matrices, which can diagonalize the mass squared matrices of CP-even sneutrino and CP-odd sneutrino. The mass matrix for chargino is diagonalized by rotation matrices $U$ and $V$.

We also deduce the vertex couplings of neutrino-slepton-chargino and neutralino-lepton-slepton,
\begin{eqnarray}
&&\mathcal{L}_{\bar{\nu}\chi^-\tilde{L}}=\bar{\nu}_i\Big((-g_2U^*_{j1}\sum_{a=1}^3U^{V*}_{ia}Z^E_{ka}+U^*_{j2}\sum_{a=1}^3U^{V*}_{ia}Y^a_lZ^E_{k(3+a)})P_L
\nonumber\\&&\hspace{1.6cm}+\sum_{a,b=1}^3Y_{\nu}^{ab}U^V_{i(3+a)}Z^E_{kb}V_{j2}P_R\Big)\chi^-_j\tilde{L}_k\;,
\\
&&\mathcal{L}_{\bar{\chi}^0l\tilde{L}}=\bar{\chi}^0_i\Big\{\Big(\frac{1}{\sqrt{2}}(g_1N^*_{i1}+g_2N^*_{i2}+g_{YX}N^*_{i5})Z^E_{kj}
-N^*_{i3}Y^j_lZ^E_{k(3+j)}\Big)P_L\nonumber\\&&\hspace{1.6cm}
-\Big[\frac{1}{\sqrt{2}}\Big(2g_1N_{i1}+(2g_{YX}+g_X)N_{i5}\Big)Z^E_{k(3+a)}+Y_{l}^jZ^E_{kj}N_{i3}\Big]P_R\Big\}l_j\tilde{L}_k\;.
\end{eqnarray}
With $Z^E$ and $N$ are rotation matrices, which can diagonalize the mass squared matrix of slepton and the mass matrix of neutralino. The mass matrix for neutrino is diagonalized by $U^V$.

Other couplings needed can be found in our previous works \cite{pro,slh}.

\section{formulation}
The Feynman amplitude can be expressed by these dimension 6 operators \cite{lepton} with the effective Lagrangian method. The dimension 8 operators be suppressed by the factor $\frac{m_{\mu}^2}{M_{SUSY}^2}$ $\sim$ ($10^{-7}$, $10^{-8}$) and can then be ignored.

These dimension 6 operators are shown below
\begin{eqnarray}
&&\mathcal{O}_1^{\mp}=\frac{1}{(4\pi)^2}\bar{l}(i\mathcal{D}\!\!\!\slash)^3\omega_{\mp}l\;,
\nonumber\\
&&\mathcal{O}_2^{\mp}=\frac{eQ_f}{(4\pi)^2}\overline{(i\mathcal{D}_{\mu}l)}\gamma^{\mu}
F\cdot\sigma\omega_{\mp}l\;,
\nonumber\\
&&\mathcal{O}_3^{\mp}=\frac{eQ_f}{(4\pi)^2}\bar{l}F\cdot\sigma\gamma^{\mu}
\omega_{\mp}(i\mathcal{D}_{\mu}l)\;,\nonumber\\
&&\mathcal{O}_4^{\mp}=\frac{eQ_f}{(4\pi)^2}\bar{l}(\partial^{\mu}F_{\mu\nu})\gamma^{\nu}
\omega_{\mp}l,\nonumber\\&&
\mathcal{O}_5^{\mp}=\frac{m_l}{(4\pi)^2}\bar{l}(i\mathcal{D}\!\!\!\slash)^2\omega_{\mp}l\;,
\nonumber\\&&\mathcal{O}_6^{\mp}=\frac{eQ_fm_l}{(4\pi)^2}\bar{l}F\cdot\sigma
\omega_{\mp}l\;.
\end{eqnarray}
Here, $\mathcal{D}_{\mu}=\partial_{\mu}+ieA_{\mu}$ and $\omega_{\mp}=\frac{1\mp\gamma_5}{2}$. $F_{{\mu\nu}}$ denotes the electromagnetic field strength, and $m_{_l}$ represents the lepton mass.

The effective Lagrangian of lepton EDM is
\begin{eqnarray}
&&{\cal L}_{_{EDM}}=\frac{-i}{2}d_l\bar{l}\sigma^{\mu\nu}\gamma_5lF_{\mu\nu}\;.
\end{eqnarray}
For Fermions EDM cannot be obtained at tree level in the fundamental interaction because it is a CP violation amplitude. Then, the one-loop diagrams should have non-zero contribution to Fermion EDM in the CP violating electroweak theory. With the relationship between the Wilson coefficients $C_{2,3,6}^{\mp}$ of the operators $\mathcal{O}_{2,3,6}^{\mp}$ \cite{lepton,edms6,edms7,edms8}, the lepton EDM obtained is shown below
\begin{eqnarray}
d_l=\frac{-2eQ_fm_l}{(4\pi)^2}\Im(C_2^{+} + C_2^{-*} +C_6^{+})\;.
\end{eqnarray}

\subsection{The one-loop corrections}
The one-loop new physics contributions to lepton EDMs come from the diagrams in FIG. 1. The one-loop contributions to lepton EDMs are obtained by calculating with the on-shell condition of external lepton. Then we simplify the analytical results.

\begin{figure}[t]
\begin{center}
\begin{minipage}[c]{0.8\textwidth}
\includegraphics[width=5.0in]{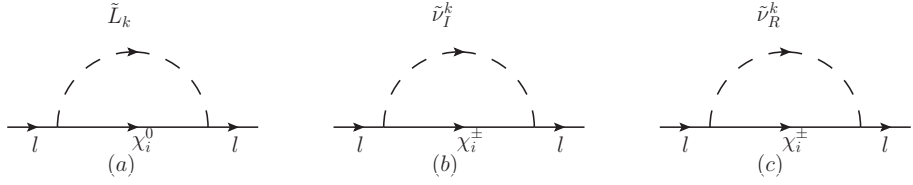}
\end{minipage}%
\caption{The one-loop self energy diagrams in the $U(1)_X$SSM.}
\end{center}
\end{figure}
The analytical results of the one-loop diagrams are shown below

1. The corrections to lepton EDMs from neutralinos and scalar leptons are
\begin{eqnarray}
&&d_{l}^{\tilde{L}\chi^{0}}=(\frac{-e}{2\Lambda})\Im\left[-\sum_{i=1}^8\sum_{j=1}^6\Big\{(A_L^*A_R)
\sqrt{x_{\chi_i^{0}}}x_{\tilde{L}_j}\frac{\partial^2 \mathcal{B}(x_{\chi_i^{0}},x_{\tilde{L}_j})}{\partial x_{\tilde{L}_j}^2}\Big\}\right]\;.
\end{eqnarray}
With $x_i=\frac{m_i^2}{\Lambda^2}$, $m_i$ represents the particle mass and $\Lambda$ denotes the new physics energy scale.
The couplings $A_R,A_L$ can be expressed as
\begin{eqnarray}
&&A_R=\frac{1}{\sqrt{2}}g_1N_{i1}^{*}Z_{j2}^{E}+\frac{1}{\sqrt{2}}g_2N_{i2}^{*}Z_{j2}^{E}+\frac{1}{\sqrt{2}}g_{YX}N_{i5}^{*}Z_{j2}^{E}
-N_{i3}^{*}Y_\mu Z_{j5}^{E}\;,\nonumber\\&&
A_L=-\frac{1}{\sqrt{2}}Z_{j5}^{E}(2g_1N_{i1}+(2g_{YX}+g_X)N_{i5})-Y_\mu^{*}Z_{j2}^EN_{i3}\;.
\end{eqnarray}
The mass matrices of scalar leptons and neutrinos can be diagonalized using the matrices $Z^{E}$ and $N$.

The specific forms of functions $\mathcal{B}(x,y)$ (using in the Eq.(11)) and $\mathcal{B}_1(x,y)$ (using in the Eqs.(14) and (16)) are shown below
\begin{eqnarray}
\mathcal{B}(x,y)=\frac{1}{16 \pi
^2}\Big(\frac{x \ln x}{y-x}+\frac{y \ln
y}{x-y}\Big)\;,~~~
\mathcal{B}_1(x,y)=(
\frac{\partial}{\partial y}+\frac{y}{2}\frac{\partial^2 }{\partial y^2})\mathcal{B}(x,y)\;.
\end{eqnarray}

2. The corrections from chargino and CP-odd scalar neutrino are
\begin{eqnarray}
&&d_{lI}^{\tilde{\nu}\chi^{\pm}}=(\frac{-e}{2\Lambda})\Im\left[\sum_{i=1}^2\sum_{j=1}^6
\Big\{-2(B_L^{*}B_R)\sqrt{x_{\chi_i^{-}}}\mathcal{B}_1(x_{\tilde{\nu}_j^{I}},x_{\chi_i^{-}})\Big\}\right]\;.
\end{eqnarray}
The couplings $B_L$ and $B_R$ can be expressed as
\begin{eqnarray}
B_L=-\frac{1}{\sqrt{2}}U_{i2}^{*}Z_{j2}^{I*}Y_\mu\;,~~~
B_R=\frac{1}{\sqrt{2}}g_2Z_{j2}^{I*}V_{i1}\;.
\end{eqnarray}

3. The corrections from chargino and CP-even scalar neutrino are
\begin{eqnarray}
&&d_{lR}^{\tilde{\nu}\chi^{\pm}}=(\frac{-e}{2\Lambda})\Im\left[\sum_{i=1}^2\sum_{j=1}^6
\Big\{-2(C_L^{*}C_R)\sqrt{x_{\chi_i^{-}}}\mathcal{B}_1(x_{\tilde{\nu}_j^{R}},x_{\chi_i^{-}})\Big\}\right]\;.
\end{eqnarray}
The couplings $C_L$ and $C_R$ can be expressed as
\begin{eqnarray}
C_L=\frac{1}{\sqrt{2}}U_{i2}^{*}Z_{j2}^{R*}Y_\mu\;,~~~
C_R=-\frac{1}{\sqrt{2}}g_2Z_{j2}^{R*}V_{i1}\;.
\end{eqnarray}
And, the $U$, $V$, $Z^{R}$ and $Z^{I}$ matrices diagonalize the corresponding particle mass matrices, which are detailed in the appendix.

So the contributions of the one-loop diagrams to lepton EDMs are
\begin{eqnarray}
&&d_l^{one-loop}=d_{l}^{\tilde{L}\chi^{0}}+d_{lI}^{\tilde{\nu}\chi^{\pm}}+d_{lR}^{\tilde{\nu}\chi^{\pm}}\;.
\end{eqnarray}

\subsection{The two-loop corrections}

 In this paper, the two-loop diagrams that we research include the Barr-Zee two-loop diagrams (FIG. 2 (a), (b), (c)) and rainbow two-loop diagrams (FIG. 2 (d), (e)), as shown below.
\begin{figure}[t]
\begin{center}
\begin{minipage}[c]{1.0\textwidth}
\includegraphics[width=6.0in]{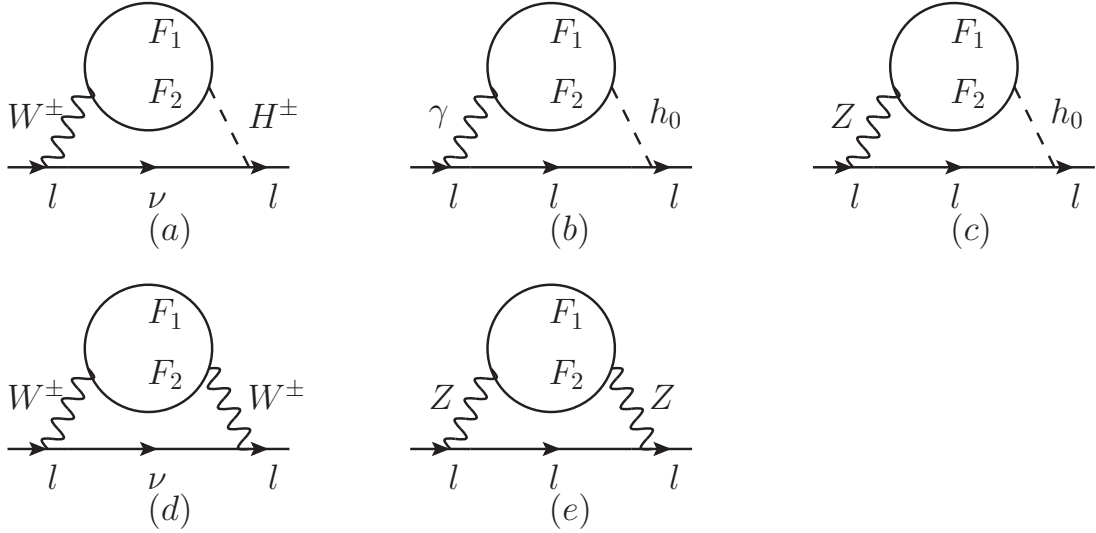}
\end{minipage}%
\caption{The two-loop Barr-Zee and rainbow type diagrams in the $U(1)_X$SSM.}
\end{center}
\end{figure}

The analytical results of the contributions from the two-loop diagrams to lepton EDMs are shown below.

The contribution from FIG. 2 (a). Under the assumption $m_F=m_{F_1}=m_{F_2}\gg m_W$, the result \cite{ffa} of simplification is
\begin{eqnarray}
&&\qquad\quad\hspace{-2cm}d_l^{WH}=\frac{-G_F m_W^2 s_W}{256\pi^4}\sum_{F_1=\chi^{\pm}}\sum_{F_2=\chi^0}\frac{H_{\bar{l}H\nu}^L}{ m_F}\Big\{\Im\Big[\Big[\frac{21}{4}-\frac{5}{18}Q_{F_1}+(3+\frac{Q_{F_1}}{3})
(\ln{m_{F_1}^2}\nonumber\\
&&\qquad\quad\hspace{-2cm}-\varrho_{1,1}(m_W^2,m_{H^\pm}^2))\Big](H_{HF_1F_2}^LH_{WF_1F_2}^L+H_{HF_1F_2}^RH_{WF_1F_2}^R)
+\Big[\frac{19-20Q_{F_1}}{9}\nonumber\\
&&\qquad\quad\hspace{-2cm}+\frac{2-4Q_{F_1}}{3}
(\ln{m_{F_1}^2}-\varrho_{1,1}(m_W^2,m_{H^\pm}^2))\Big](H_{HF_1F_2}^LH_{WF_1F_2}^R+H_{HF_1F_2}^RH_{WF_1F_2}^L)
\nonumber\\
&&\qquad\quad\hspace{-2cm}+\Big[\hspace{-0.1cm}-\hspace{-0.1cm}\frac{16}{9}\hspace{-0.1cm}-\hspace{-0.1cm}\frac{2\hspace{-0.1cm}+\hspace{-0.1cm}6Q_{F_1}}{3}
(\ln{m_{F_1}^2}\hspace{-0.1cm}-\hspace{-0.1cm}\varrho_{1,1}(m_W^2,m_{H^\pm}^2))\Big](H_{HF_1F_2}^LH_{WF_1F_2}^L\hspace{-0.1cm}-\hspace{-0.1cm}H_{HF_1F_2}^RH_{WF_1F_2}^R)
\nonumber\\
&&\qquad\quad\hspace{-2cm}+\Big[\hspace{-0.1cm}-\hspace{-0.1cm}\frac{2Q_{F_1}}{9}\hspace{-0.1cm}-\hspace{-0.1cm}\frac{6\hspace{-0.1cm}-\hspace{-0.1cm}2Q_{F_1}}{3}
(\ln{m_{F_1}^2}\hspace{-0.1cm}-\hspace{-0.1cm}\varrho_{1,1}(m_W^2,m_{H^\pm}^2))\Big](H_{HF_1F_2}^LH_{WF_1F_2}^R\hspace{-0.1cm}-\hspace{-0.1cm}H_{HF_1F_2}^RH_{WF_1F_2}^L)\Big]\Big\}\;.
\end{eqnarray}
Here, $\varrho_{1,1}(x,y)=\frac{x\ln x-y\ln y}{x-y}$. $H_{HF_1F_2}^{L,R}$ and $H_{WF_1F_2}^{L,R}$ denote the corresponding couplings coefficients. Please see the Ref.\cite{slh} for their concrete forms.

Under the assumption $m_F=m_{F_1}=m_{F_2}\gg m_{h_0}$, the reduced form of the contribution to lepton EDM from FIG. 2(b) is
\begin{eqnarray}
&&d_l^{\gamma h_0}=\frac{-eG_FQ_fQ_{F_1}m_W^2s_W^2}{32\pi^4}\sum_{F_1=F_2=\chi^\pm}\Big\{\Im\Big[\frac{1}{m_{F_1}}
(H_{h_0F_1F_2}^L)[1+\ln\frac{m_{F_1}^2}{m_{h_0}^2}]\Big]\Big\}\;.
\end{eqnarray}

And, the simplified form from FIG. 2(c) is given below
\begin{eqnarray}
&&d_l^{Zh_0}=\frac{-\sqrt{2}e}{1024\pi^4}\sum_{F_1=F_2=\chi^{\pm},\chi^0}
\Big\{\frac{H_{h_0l\bar{l}}}{m_{F_1}}\Big[\varrho_{1,1}(m_Z^2,m_{h_0}^2)-\ln{m_{F_1}^2}-1\Big]
\nonumber\\&&\qquad\quad\times\Im[(H^L_{Zll}-H^R_{Zll})(H_{h_0F_1F_2}^LH_{ZF_1F_2}^L+H_{h_0F_1F_2}^RH_{ZF_1F_2}^R)]\Big\}\;.
\end{eqnarray}
With $Q_f$ represents the electric charge of the external lepton $m_\mu$. $Q_{F_1}$ and $Q_{F_2}$ denote the electric charges of the internal charginos.

With the assumption $m_F=m_{F_1}=m_{F_2}\gg m_W\sim m_Z$, the reduced form of the contribution to lepton EDM from FIG. 2(d) is
\begin{eqnarray}
&&d_l^{WW}=\frac{-eG_F m_l}{384\sqrt{2}\pi^4}\sum_{F_1=\chi^{\pm}}\sum_{F_2=\chi^0}\left\{\Im[11(H_{WF_1F_2}^{R*}H_{WF_1F_2}^L)]\right\}\;.
\end{eqnarray}

We simplify the tedious two-loop results to the order $\frac{m_\mu^2}{M_Z^2}$ $\sim$ $10^{-6}$ or $\frac{m_\mu^2}{m_{SUSY}^2}$ under the assumption $m_F = m_{F1} = m_{F2} \gg m_W \sim m_Z$, and get the simplified form of FIG. 2(e) as follows

\begin{eqnarray}
&&d_{l}^{ZZ}=
\frac{eQ_{F_1}m_l}{2048\Lambda^2\pi^4}\sum_{F_1=F_2=\chi^\pm}\Big\{
\Im\Big[(H^L_{ZF_1F_2}H^R_{ZF_1F_2})\Big(|H^L_{Zll}|^2
+|H^R_{Zll}|^2\Big)[\frac{-6 \log x_Z+6 \log x_F+4}{9 x_F}]\nonumber\\&&
+\Big(|H^L_{ZF_1F_2}|^2+|H^R_{ZF_1F_2}|^2\Big)H^L_{Zll}H^R_{Zll}
[16\frac{(\log x_F-\log x_Z) (\log x_F+2)+2}{x_Z}]\Big]\Big\}\;.
\end{eqnarray}

The contributions to lepton EDMs from the researched two-loop diagrams are
\begin{eqnarray}
&&d_l^{two-loop}=d_l^{WH}+d_l^{\gamma h_0}+d_l^{Z h_0}+d_l^{WW}+d_{l}^{ZZ}\;.
\end{eqnarray}

At two-loop level, the contributions to lepton EDMs can be summarized as
\begin{eqnarray}
&&d_l^{total}=d_l^{one-loop}+d_l^{two-loop}\;.
\end{eqnarray}

\section{the numerical results}
For the numerical discussion, we consider the following experimental limitations. The lightest CP-even higgs mass is considered as an input parameter, which is $m_{h^0}$ $\approx$ 125.1 GeV \cite{hmass1, hmass2}. And $h^0$ decays are $h^0 \rightarrow \gamma + \gamma$, $h^0 \rightarrow Z + Z$ and $h^0 \rightarrow \gamma + \gamma$ \cite{dec}. Experimental constraints on the masses of the new particles are also considered. The LHC experiments have more stringent mass constraints on $Z^{\prime}$ boson. To satisfy the experimental constraint, we take the parameter $M_{Z^{\prime}}$ greater than 5.1 TeV \cite{51}, which is heavier than the previous mass limit. The ratio of $M_{Z^{\prime}}$ to its gauge coupling $g_X$, $(\frac{M_{Z^{\prime}}}{g_X})$ should be not less than 6 TeV at $99\%$ C.L. \cite{cl1, cl2}. Considering the constraints from the LHC, we set the $\tan{\beta_\eta} < 1.5$ \cite{BT}. Since $M_{Z^{\prime}}$ has a large mass, the contribution of $Z^{\prime}$ at the amplitude level is very small, so the contribution of $Z^{\prime}$ is ignored.
Considering the experimental limitation of lepton EDMs, we adjust the parameters. In this section, we research and discuss lepton ($e, \mu, \tau$)EDMs respectively.

The parameters used in $U(1)_X$SSM are given below:
\begin{eqnarray}
&&g_X=0.33,~g_{YX}=0.2,~\lambda_C=-0.1,~\kappa=0.1,~T_{\lambda_H}=1.0~{\rm TeV},~T_{\kappa}=1.0~{\rm TeV},
\nonumber\\&&\tan{\beta_\eta}=1.05,~v_{\eta}=15\times\cos{\beta_\eta}~{\rm TeV},~v_{\bar{\eta}}=15\times\sin{\beta_\eta}~{\rm TeV},
~B_\mu=8~{\rm TeV^2},
\nonumber\\&&m_S^2=8~{\rm TeV^2},~T_{\lambda_C}=150~{\rm GeV}, ~T_{E11}=T_{E22}=T_{E33}=0.1~{\rm TeV},
\nonumber\\&&M_{\nu11}=M_{\nu22}=M_{\nu33}=6~{\rm TeV^2}, ~Y_{X11}=Y_{X22}=Y_{X33}=0.04, \nonumber\\&&B_S=8~{\rm TeV^2}, ~\lambda_H=0.1, ~l_W=8~{\rm TeV^2}, ~T_{X11}=T_{X22}=T_{X33}=10~{\rm GeV}.
\end{eqnarray}
$\theta_1$, $\theta_2$ and $\theta_\mu$ are the CP violating phases of the parameters $m_1$, $m_2$ and $\mu$. We take into account three new CP violating parameters with the phases $\theta_{BL}$, $\theta_{BB^{\prime}}$ and $\theta_S$.
\begin{eqnarray}
&&m_1=M_1*e^{i*\theta_{1}}, ~m_2=M_2*e^{i*\theta_{2}}, ~\mu=mu*e^{i*\theta_{\mu}},
\nonumber\\&&m_{BL}=M_{BL}*e^{i*\theta_{BL}}, ~m_{{BB}^\prime}=M_{{BB}^\prime}*e^{i*\theta_{BB^{\prime}}}, ~m_S=M_S*e^{i*\theta_S}.
\end{eqnarray}

In order to facilitate the following discussion, we have made some simplifications:
\begin{eqnarray}
&&M_L=M_{L11}=M_{L22}=M_{L33}, ~~~M_E=M_{E11}=M_{E22}=M_{E33},
\nonumber\\&&T_E=T_{E11}=T_{E22}=T_{E33}.
\end{eqnarray}
\subsection{the e EDM}
\begin{figure}[t]
\begin{center}
\begin{minipage}[c]{0.48\textwidth}
\includegraphics[width=2.9in]{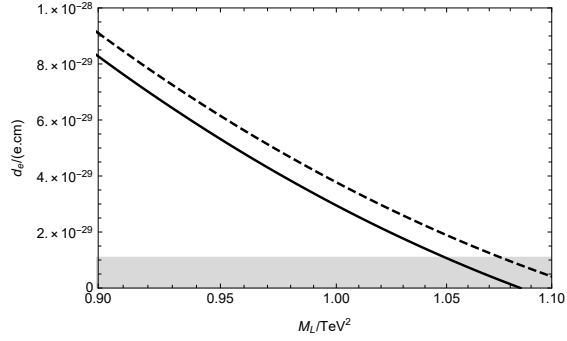}
\end{minipage}%
\caption{With $\theta_1$ = $\theta_2$ = $\theta_\mu$ = $\theta_{BB^{\prime}}$ = $\theta_S$ = 0, and $\theta_{BL}$ = $\frac{\pi}{4}$, the contributions to electron EDM varying with $M_L$ are plotted by the solid line, dashed line respectively corresponding to $M_1$ = ($700,800)~{\rm GeV}$.}\label{DEBL}
\end{center}
\end{figure}

At the beginning, we discussed the EDM of electron, because its experimental upper limit is very strict. The CP violating phases $\theta_1$, $\theta_2$ ,$\theta_\mu$, $\theta_{BL}$, $\theta_{BB^{\prime}}$ and $\theta_S$, also including other parameters have a certain impact on the electron EDM.
Now, supposing $\theta_1$ = $\theta_2$ = $\theta_\mu$ = $\theta_{BB^{\prime}}$ = $\theta_S$ = 0, and setting $\tan{\beta}=5$, $M_2=500~
{\rm GeV}$, $mu=500~{\rm GeV}$, $M_{BL}= 1800 ~{\rm GeV}$, $M_{BB^{\prime}}= 700 ~{\rm GeV}$, $M_S= 2400 ~{\rm GeV}$, $M_L=1.1 ~{\rm TeV}$, $M_E=1.0 ~{\rm TeV}$. We study the influence of $\theta_{BL}$ on electron EDM. $M_{BL}$ is related to neutralino mass matrix.
In FIG. \ref{DEBL}, we plot the solid line and dashed line versus $M_L$ ($0.9\sim1.1 ~{\rm TeV^2}$) corresponding to $M_1$ = $700,800 ~{\rm GeV}$. We can see that these two lines are subtractive functions, and $\theta_{BL}$ has influence on $|d_e|$. The relationship between $d_e$ and $M_L$ is not a simple linear relation, its change curve is like $M_L^{-2}$. The shaded part of the figure indicates that all these parameters are within reasonable parameters and conform to experimental limits.

\begin{figure}[t]
\begin{center}
\begin{minipage}[c]{0.48\textwidth}
\includegraphics[width=2.9in]{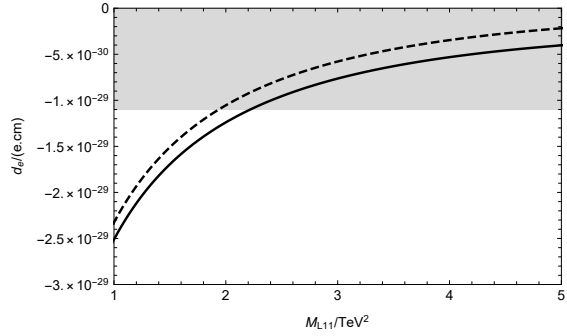}
\end{minipage}%
\caption{With $\theta_1$ = $\theta_2$ = $\theta_\mu$ = $\theta_{BB^{\prime}}$ = $\theta_{BL}$ = 0, and $\theta_S$ = $\frac{\pi}{4}$, the contributions to electron EDM varying
with $M_{L11}$ are plotted by the solid line, dashed line respectively corresponding to $M_{L33}$ = $(1, 0.9)~{\rm TeV^2}$.}\label{DES}
\end{center}
\end{figure}

Setting $\theta_1$ = $\theta_2$ = $\theta_\mu$ = $\theta_{BB^{\prime}}$ = $\theta_{BL}$ = 0, $\tan{\beta}=5$, $M_1= 700~{\rm GeV}$, $M_2=2000~
{\rm GeV}$, $mu=500~{\rm GeV}$, $M_{BL}= 1600 ~{\rm GeV}$, $M_{BB^{\prime}}= 800 ~{\rm GeV}$, $M_S= -800 ~{\rm GeV}$, $M_{L22}=1.0 ~{\rm TeV^2}$, $M_E=1.0 ~{\rm TeV^2}$, we consider the impact of $\theta_S$ on the electron EDM. $M_S$ is related to the mass matrices of neutralino and scalar lepton. In FIG. \ref{DES}, $M_{L11}$ varies from $0.5$ to $5.0$ $~{\rm TeV^2}$, and when $M_{L11}$ $>$ $2.0 ~{\rm TeV^2}$, the numerical results of $|d_e|$ conform to the experimental limits.

\begin{figure}[t]
\begin{center}
\begin{minipage}[c]{0.48\textwidth}
\includegraphics[width=2.9in]{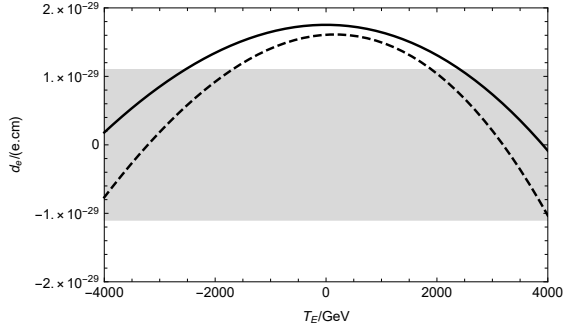}
\end{minipage}%
\caption{With $\theta_1$ = $\theta_2$ = $\theta_\mu$ = $\theta_S$ = $\theta_{BL}$ = 0, and $\theta_{BB^{\prime}}$ = $\frac{\pi}{3}$, the contributions to electron EDM varying with $T_E$ are plotted by the solid line, dashed line respectively corresponding to $M_{E11}$ = $(0.5,1.0)~{\rm TeV^2}$.}\label{DEBBp}
\end{center}
\end{figure}

$\theta_{BB^{\prime}}$ is the new CP violating phase of the lepton neutrino mass matrix. So, it make new physical contribution to the lepton EDM. With $\theta_1$ = $\theta_2$ = $\theta_\mu$ = $\theta_S$ = $\theta_{BL}$ = 0, the contributions to muon EDM varying with $T_E$ are plotted by the solid line and dashed line respectively corresponding to $M_{E11}$ = 0.5 and 1.0$~{\rm TeV^2}$. In this part, we set $\tan{\beta}=5$, $M_1= 700~{\rm GeV}$, $M_2=2000~
{\rm GeV}$, $mu=500~{\rm GeV}$, $M_{BL}= 1800 ~{\rm GeV}$, $M_{BB^{\prime}}= 700 ~{\rm GeV}$, $M_S= 2400 ~{\rm GeV}$, $M_L=1.0 ~{\rm TeV^2}$, $M_E=0.5 ~{\rm TeV^2}$. In FIG. \ref{DEBBp}, the two lines are shaped like parabolas. And most of the numerical results are within experimental limits.

\begin{figure}[t]
\begin{center}
\begin{minipage}[c]{0.48\textwidth}
\includegraphics[width=2.9in]{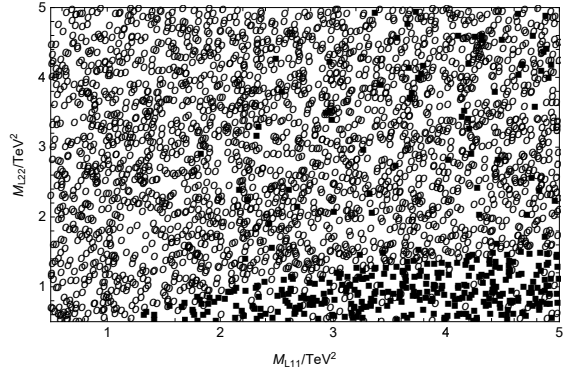}
\end{minipage}%
\caption{With $\theta_1$ = $\theta_2$ = $\theta_\mu$ = $\theta_{BB^{\prime}}$ = $\theta_{BL}$ = 0, and $\theta_S$ = $\frac{\pi}{4}$, $|d_e|$ is in the plane of $M_{L11}$ versus $M_{L22}$, ``$\blacksquare$'' represents $|d_e| < 1.1 \times 10^{-29}$ e.cm, ``$\circ$'' represents $|d_e| \geqslant 1.1 \times 10^{-29}$ e.cm.} \label{DESSJ}
\end{center}
\end{figure}

We select these parameters $M_{L11}(0.5\thicksim5.0 ~{\rm TeV^2})$, $M_{L22}(0.5\thicksim5.0 ~{\rm TeV^2})$, $M_{L33}(0.5\thicksim5.0 ~{\rm TeV^2})$, $T_E(-3000\thicksim3000 ~{\rm GeV})$, $M_E(0.5\thicksim5.0 ~{\rm TeV^2})$, and randomly scatter points.
With $\theta_1$ = $\theta_2$ = $\theta_\mu$ = $\theta_{BB^{\prime}}$ = $\theta_{BL}$ = 0, and $\theta_S$ = $\frac{\pi}{4}$. We plot $|d_e|$ in the plane of $M_{L11}$ versus $M_{L22}$ in Fig. \ref{DESSJ}. ``$\blacksquare$'' represents $|d_e| < 1.1 \times 10^{-29}$ e.cm, ``$\circ$'' represents $|d_e| \geqslant 1.1 \times 10^{-29}$ e.cm. In Fig. \ref{DESSJ}, We can see that there is a clear stratification. When $M_{L11}$ $>$ 1.0 $~{\rm TeV^2}$, $M_{L22}$ is in the vicinity of 1.4 $~{\rm TeV^2}$, $|d_e|$ is within the experimental limit. This can show that $M_{L11}$ is a sensitive parameters and $M_{L22}$ is a less sensitive parameter.
\subsection{the $\mu$ EDM}
\begin{figure}[t]
\begin{center}
\begin{minipage}[c]{0.48\textwidth}
\includegraphics[width=2.9in]{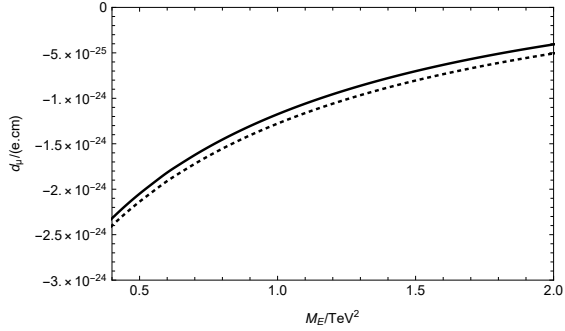}
\end{minipage}%
\caption{With $\theta_1$ = $\theta_2$ = $\theta_\mu$ = $\theta_{BB^{\prime}}$ = $\theta_{BL}$ = 0, and $\theta_S$ = $\frac{\pi}{3}$, the contributions to muon EDM varying with $M_E$ are plotted by the solid line, dashed line respectively corresponding to $M_{BL}$ = $(1200,1500)~{\rm GeV}$.} \label{DMUS}
\end{center}
\end{figure}

In this section, the muon EDM is numerically studied. In FIG. \ref{DMUS}, setting $\theta_1$ = $\theta_\mu$ = $\theta_{BB^{\prime}}$ = $\theta_2$ = $\theta_{BL}$ = 0, and setting $\tan{\beta}=6$, $M_1= 1450~{\rm GeV}$, $M_2=2000~
{\rm GeV}$, $mu=500~{\rm GeV}$, $M_{BB^{\prime}}= 800 ~{\rm GeV}$, $M_S= -800 ~{\rm GeV}$, $M_L=1.0 ~{\rm TeV^2}$, $M_E=0.5 ~{\rm TeV^2}$. We study the influence of $\theta_S$ on the muon EDM. These solid line, dashed line correspond to $M_{BL}$ ($1200, 1500 ~{\rm GeV}$). We can see that the numerical result of the muon EDM increases as $M_E$ increases. The $\theta_S$ has great influence on the numerical results, because of that $M_S$ is related to the mass matrices of neutralino and charge Higgs.

\begin{figure}[t]
\begin{center}
\begin{minipage}[c]{0.48\textwidth}
\includegraphics[width=2.9in]{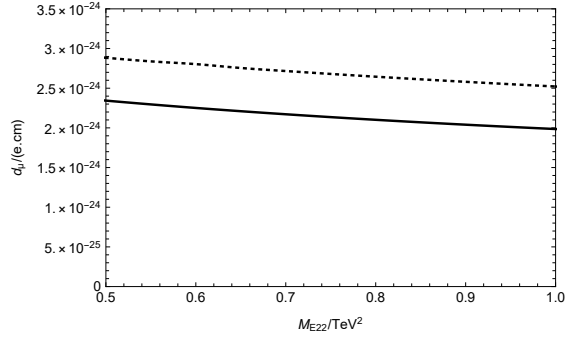}
\end{minipage}%
\caption{With $\theta_1$ = $\theta_2$ = $\theta_\mu$ = $\theta_S$ = $\theta_{BL}$ = 0, and $\theta_{BB^{\prime}}$ = $\frac{\pi}{6}$, the contributions to muon EDM varying with $M_{E22}$ are plotted by the solid line, dashed line respectively corresponding to $\tan\beta$ = ($5,6$).} \label{DMUBBP}
\end{center}
\end{figure}

$\theta_{BB^{\prime}}$ is the new CP violating phase of the neutralino mass matrix. So, it makes new physical contribution to the lepton EDMs. With $\theta_1$ = $\theta_2$ = $\theta_\mu$ = $\theta_S$ = $\theta_{BL}$ = 0, the contributions to muon EDM varying with $M_{E22}$ are plotted by the solid line and dashed line respectively corresponding to $\tan\beta$ = (5,~6). In this part, we set $M_1= 1450 ~{\rm GeV}$, $M_2= 800 ~{\rm GeV}$, $mu=500~{\rm GeV}$, $M_{BL}= 1600 ~{\rm GeV}$, $M_{BB^{\prime}}= 800 ~{\rm GeV}$, $M_S= -800 ~{\rm GeV}$, $M_L=1.0 ~{\rm TeV^2}$, $M_E=0.5 ~{\rm TeV^2}$. In FIG. \ref{DMUBBP}, as $M_{E22}$ increasing, the numerical result decreases slowly, and the shapes of the two lines are similar.

\begin{figure}[t]
\begin{center}
\begin{minipage}[c]{0.48\textwidth}
\includegraphics[width=2.9in]{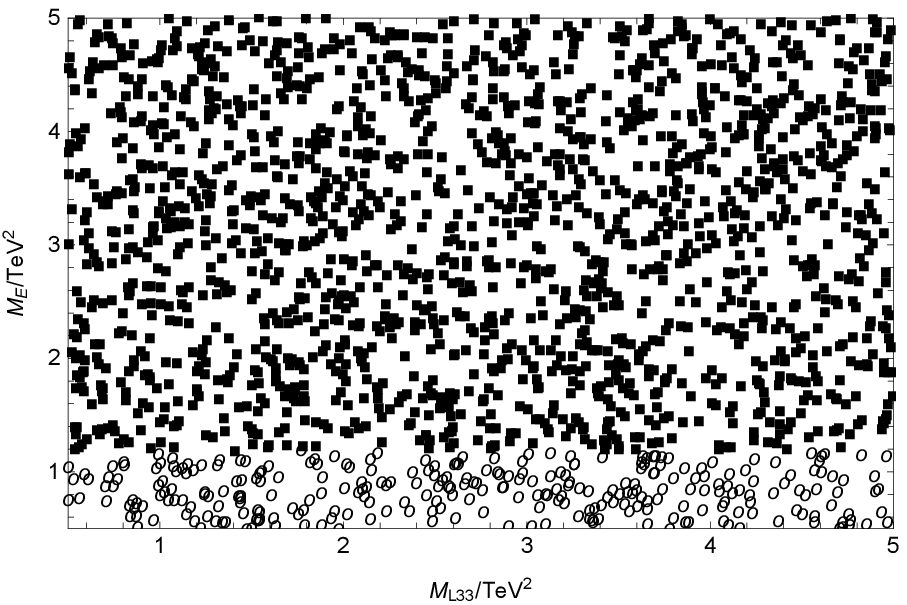}
\end{minipage}%
\caption{With $\theta_1$ = $\theta_2$ = $\theta_\mu$ = $\theta_{BB^{\prime}}$ = $\theta_{BL}$ = 0, and $\theta_S$ = $\frac{\pi}{4}$, $|d_{\mu}|$ is in the plane of $M_{L33}$ versus $M_{E}$, ``$\blacksquare$'' represents $|d_\mu|$ $<$ $1\times10^{-24}$ e.cm, ``$\circ$'' represents $|d_\mu|$ $\geqslant$ $1\times10^{-24}$ e.cm.} \label{DMUSSJ}
\end{center}
\end{figure}

We choose these parameters $M_{L11}(0.5\thicksim5.0 ~{\rm TeV^2})$, $M_{L22}(0.5\thicksim5.0 ~{\rm TeV^2})$, $M_{L33}(0.5\thicksim5.0 ~{\rm TeV^2})$, $T_E(-3000\thicksim3000 ~{\rm GeV})$, $M_E(0.5\thicksim5.0 ~{\rm TeV^2})$, and randomly scatter points.
With $\theta_1$ = $\theta_2$ = $\theta_\mu$ = $\theta_{BB^{\prime}}$ = $\theta_{BL}$ = 0, and $\theta_S$ = $\frac{\pi}{4}$, we study $|d_\mu|$ in the plane of $M_{L33}$ versus $M_E$. In FIG. \ref{DMUSSJ}, ``$\blacksquare$'' represents $|d_\mu|$ $<$ $1\times10^{-24}$ e.cm, ``$\circ$'' represents $|d_\mu|$ $\geqslant$ $1\times10^{-24}$ e.cm. Delamination occurs when $M_E$ = $1.1 ~{\rm TeV^2}$, and the stratification is obvious. This can show that $M_{E}$  is a sensitive parameter and $M_{L33}$ is an insensitive parameter. These parameters are in a reasonable parameter space.

\subsection{the $\tau$ EDM}
\begin{figure}[t]
\begin{center}
\begin{minipage}[c]{0.48\textwidth}
\includegraphics[width=2.9in]{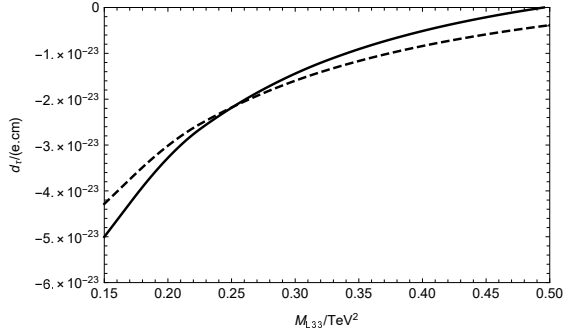}
\end{minipage}%
\caption{With $\theta_1$ = $\theta_2$ = $\theta_\mu$ = $\theta_{BB^{\prime}}$ = $\theta_{BL}$ = 0, and $\theta_S$ = $\frac{\pi}{5}$, the contributions to tau EDM varying with $M_{L33}$ are plotted by the solid line, dashed line respectively corresponding to $M_2$ = $(400,500)~{\rm GeV}$.} \label{DTAUS}
\end{center}
\end{figure}

At present, the experimental upper bound of tau EDM is $|d^{exp}_{\tau}|$ $<$ $1.1 \times 10^{-17}$ e.cm, and it is largest one among bounds of the lepton EDMs. So, we study the tau EDM in this subsection. Setting $\tan{\beta}=6$, $M_1= 750 ~{\rm GeV}$, $mu=650 ~{\rm GeV}$, $M_{BL}=1800 ~{\rm GeV}$, $M_{BB^{\prime}}= 700 ~{\rm GeV}$, $M_S= 1400 ~{\rm GeV}$, $M_L=1.0 ~{\rm TeV^2}$, $M_E=1.0 ~{\rm TeV^2}$, and setting $\theta_1$ = $\theta_2$ = $\theta_\mu$ = $\theta_{BB^{\prime}}$ = $\theta_{BL}$ = 0, and $\theta_S$ = $\frac{\pi}{5}$, we study the influence of $M_{L33}$ on $|d_\tau|$. In FIG. \ref{DTAUS}, the solid line and dashed line respectively correspond to $M_2$ = $(400,500 ~{\rm GeV})$ and their numerical results are all in the negative part. The two lines are increasing functions of $M_{L33}$, and $\theta_S$ has more obvious influence on numerical result of $|d_{\tau}|$. The maximum value of two lines can reach $5.0 \times 10^{-23}$ e.cm, and this value is 6 orders of magnitude smaller than the upper limit of the experiment.

\begin{figure}[t]
\begin{center}
\begin{minipage}[c]{0.48\textwidth}
\includegraphics[width=2.9in]{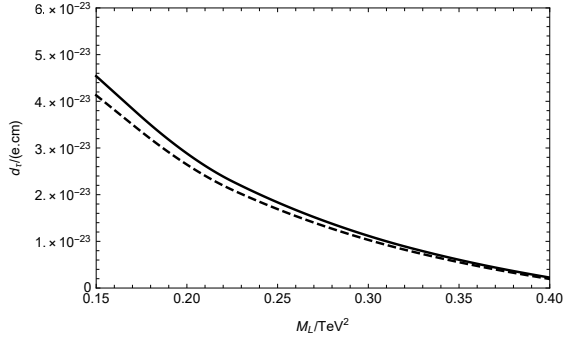}
\end{minipage}%
\caption{With $\theta_1$ = $\theta_2$ = $\theta_\mu$ =  $\theta_S$ = $\theta_{BB^{\prime}}$ = 0, and $\theta_{BL}$ = $\frac{\pi}{6}$, the contributions to tau EDM varying with $M_L$ are plotted by the solid line, dashed line respectively corresponding to $mu$ = $(650,750)~{\rm GeV}$.} \label{DTAUBL}
\end{center}
\end{figure}

$\theta_{BL}$ is the new CP violating phase of $M_{BL}$ in the neutralino mass matrix. Setting $\tan{\beta}=6$, $M_1= 750 ~{\rm GeV}$, $M_2= 400 ~{\rm GeV}$, $M_{BL}=1800 ~{\rm GeV}$, $M_{BB^{\prime}}= 700 ~{\rm GeV}$, $M_S= 1400 ~{\rm GeV}$, $M_E=1.0 ~{\rm TeV^2}$, $\theta_1$ = $\theta_2$ = $\theta_\mu$ = $\theta_{BB^{\prime}}$ = $\theta_S$ = 0, and $\theta_{BL}$ = $\frac{\pi}{6}$, the contributions to tau EDM varying with $M_L$ are plotted by the solid line and dashed line respectively corresponding to $mu$ = $(650,750 ~{\rm GeV}$). In FIG. \ref{DTAUBL}, we can see that $|d_\tau|$ decreases with the increase of $M_L$. The maximum value of these two lines can reach $|d_\tau|$ = $4.5 \times 10^{-23}$ e.cm.

\begin{figure}[t]
\begin{center}
\begin{minipage}[c]{0.48\textwidth}
\includegraphics[width=2.9in]{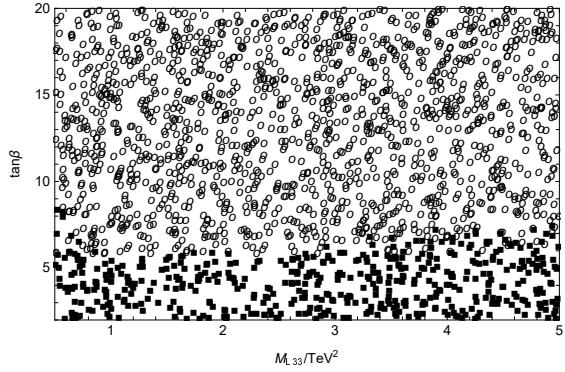}
\end{minipage}%
\caption{With $\theta_1$ = $\theta_2$ = $\theta_\mu$ = $\theta_{BB^{\prime}}$ = $\theta_S$ = 0, and $\theta_{BL}$ = $\frac{\pi}{6}$, $|d_{\tau}|$ is in the plane of $M_{L33}$ versus $\tan{\beta}$, ``$\blacksquare$'' represents $|d_\tau|$ $<$ $1 \times 10^{-23}$ e.cm, ``$\circ$'' represents $|d_\tau|$ $\geqslant$ $1 \times 10^{-23}$ e.cm.} \label{DTAUSJ}
\end{center}
\end{figure}

We select these parameters $M_{L11}(0.5\thicksim5.0 ~{\rm TeV^2})$, $M_{L22}(0.5\thicksim5.0 ~{\rm TeV^2})$, $M_{L33}(0.5\thicksim5.0 ~{\rm TeV^2})$, $T_E(-3000\thicksim3000 ~{\rm GeV})$, $\tan{\beta}(2\thicksim20)$, and randomly scatter points.
In Fig. \ref{DTAUSJ}, we study $|d_\tau|$ in the plane of $M_{L33}$ and $\tan{\beta}$ to see their influence. The varying regions of $M_{L33}$ and $\tan{\beta}$ are in the range $(0.5\thicksim5 ~\rm TeV^2)$ and $(2\thicksim20)$ respectively.``$\blacksquare$'' represents $|d_\tau|$ $<$ $1 \times 10^{-23}$ e.cm, ``$\circ$'' represents $|d_\tau|$ $\geqslant$ $1 \times 10^{-23}$ e.cm. When $\tan{\beta}$ = 6, stratification occurs, and the stratification is more obvious. This indicates that $\tan{\beta}$ is a sensitive parameter.
\section{discussion and conclusion}
In the $U(1)_X$SSM, we calculate and analyze the one-loop and two-loop contributions to the lepton ($e,\mu,\tau$) EDMs. The effects of the CP violating phases $\theta_1$, $\theta_2$, $\theta_\mu$, $\theta_{BB^{\prime}}$, $\theta_S$, $\theta_{BL}$ to the lepton EDMs are researched. Among them, $\theta_{BB^{\prime}}$, $\theta_S$, $\theta_{BL}$ are all newly introduced ones. The experimental upper limit of electron EDM is $|d^{exp}_e|$ $<$ $1.1 \times 10^{-29}$ e.cm, which gives strict restrictions on the $U(1)_X$SSM parameter space. In the our used parameter space, the numerical result of $|d_e|$ can be controlled below the experimental limit. In our study, the largest numerical results of $\mu$ EDM and $\tau$ EDM are about $2.8 \times 10^{-24}$ e.cm and $5.0 \times 10^{-23}$ e.cm  respectively. They are all in a reasonable parameter space and do not exceed the upper limit of the experiment. For the corrections of lepton EDMs, the one-loop contributions are dominant. As for the contributions of one-loop and two-loop to EDMs, their relative size $(d_l^{two-loop} /d_l^{one-loop})$ are about $5\%\thicksim15\%$ after numerical calculation.

Our numerical results mainly obey the rule $d_e / d_\mu / d_\tau$ $\thicksim$ $m_e / m_\mu / m_\tau$. In FIG. \ref{DEBL}, when $\theta_{BL}$ = $\frac{\pi}{4}$, $M_L$ has a more obvious impact on electron EDM, and the influence of $\theta_{BL}$ on electron EDM is also more obvious.
In addition, the influences of the CP-violating phases $\theta_{S}$ and $\theta_{BB^{\prime}}$ on lepton EDMs are also obvious. In FIG. \ref{DMUS}, when $\theta_{S}$ = $\frac{\pi}{3}$, the value of the muon EDM increases as $M_E$ increases (the numerical results are all negative), The $\theta_{S}$ has great influence on the numerical results, because of that $M_S$ is related to the mass matrices of neutralino and charge Higgs. In FIG. \ref{DMUBBP}, when $\theta_{BB^{\prime}}$ = $\frac{\pi}{6}$, the two lines (solid line, dashed line) are about the decreasing function of $M_{E22}$. The above parameters ($M_L$,~$M_E$) are all elements on the diagonal of the mass matrix, so their corresponding results are all decoupled, such as FIG. \ref{DEBL}, FIG. \ref{DES}, FIG. \ref{DMUS}, FIG. \ref{DMUBBP}, FIG. \ref{DTAUS}, FIG. \ref{DTAUBL}. In FIG. 12, We can get that $|d_\tau|$ increases with the increase of $\tan\beta$. If we use the method of mass insertion \cite{massi} to analyze the results, it is intuitive to find that $\tan\beta$ is proportional to lepton EDMs. We have also performed some random spot operations on lepton EDMs. The randomly scattered pictures have obvious stratification, also help us to find a reasonable parameter space. As the accuracy of technology improves, lepton EDMs may be detected in the near future.

\section{acknowledgments}
This work is supported by National Natural Science Foundation of China(NNSFC)(Nos. 11535002, 11705045), Natural Science Foundation of Hebei Province (A2020201002) and the youth top-notch talent support program of the Hebei Province.

\appendix
\begin{center}
\Large{{\bf Appendix }}
\end{center}
The mass matrix for slepton with the basis $(\tilde{e}_L,\tilde{e}_R)$
\begin{eqnarray}
m_{\tilde{e}}^2=
\left({\begin{array}{*{20}{c}}
m_{\tilde{e}_{L}\tilde{e}_{L}^*} & \frac{1}{2}(\sqrt{2}v_dT_e^\dag - v_u(\lambda_Hv_S + \sqrt{2}\mu)Y_e^\dag) \\
\frac{1}{2}(\sqrt{2}v_dT_e - v_uY_e(\sqrt{2}\mu^* + v_S\lambda_H^*)) & m_{\tilde{e}_{R}\tilde{e}_{R}^*} \\
\end{array}}
\right)\;,
\end{eqnarray}
\begin{eqnarray}
&& m_{\tilde{e}_{L}\tilde{e}_{L}^*}=m_{\tilde{l}}^{2}+\frac{1}{8}\Big((g_{1}^{2}+g_{YX}^{2}+g_{YX}g_{X}-g_{2}^{2})(v_d^2-v_u^2)
+2g_{YX}g_{X}(v_\eta^2-v_{\bar{\eta}}^2)\Big)+\frac{1}{2}v_d^2Y_e^{\dag} Y_e,
\nonumber\\&& m_{\tilde{e}_{R}\tilde{e}_{R}^*}=m_e^2-\frac{1}{8}\Big([2(g_1^2+g_{YX})+3g_{YX}g_X+g_X^2](v_d^2-v_u^2)
+(4g_{YX}g_X+2g_X^2)(v_\eta^2-v_{\bar{\eta}}^2)\Big)
\nonumber\\&&\hspace{1.8cm}+\frac{1}{2}v_d^2Y_eY_e^{\dag}\;.
\end{eqnarray}
This matrix is diagonalized by $Z^E$
\begin{eqnarray}
Z^Em_{\tilde{e}}^2Z^{E,\dag} = m_{2,\tilde{e}}^{dia}\;.
\end{eqnarray}

The mass matrix for CP-even sneutrino $({\phi}_{l}, {\phi}_{r})$ reads
\begin{eqnarray}
m^2_{\tilde{\nu}^R} = \left(
\begin{array}{cc}
m_{{\phi}_{l}{\phi}_{l}} &m^T_{{\phi}_{r}{\phi}_{l}}\\
m_{{\phi}_{l}{\phi}_{r}} &m_{{\phi}_{r}{\phi}_{r}}\end{array}
\right)\;,
\end{eqnarray}
\begin{eqnarray}
&&m_{{\phi}_{l}{\phi}_{l}}= \frac{1}{8} \Big((g_{1}^{2} + g_{Y X}^{2} + g_{2}^{2}+ g_{Y X} g_{X})( v_{d}^{2}- v_{u}^{2})
+  g_{Y X} g_{X}(2 v_{\eta}^{2}-2 v_{\bar{\eta}}^{2})\Big)
\nonumber\\&&\hspace{1.8cm}+\frac{1}{2} v_{u}^{2}{Y_{\nu}^{T}  Y_\nu}  + m_{\tilde{L}}^2\;,
 \\&&m_{{\phi}_{l}{\phi}_{r}} = \frac{1}{\sqrt{2} } v_uT_\nu  +  v_u v_{\bar{\eta}} {Y_X  Y_\nu}
  - \frac{1}{2}v_d ({\lambda}_{H}v_S  + \sqrt{2} \mu )Y_\nu\;,\\&&
m_{{\phi}_{r}{\phi}_{r}}= \frac{1}{8} \Big((g_{Y X} g_{X}+g_{X}^{2})(v_{d}^{2}- v_{u}^{2})
+2g_{X}^{2}(v_{\eta}^{2}- v_{\bar{\eta}}^{2})\Big) + v_{\eta} v_S Y_X {\lambda}_{C}\nonumber \\&&\hspace{1.8cm}
 +m_{\tilde{\nu}}^2 + \frac{1}{2} v_{u}^{2}|Y_\nu|^2+  v_{\bar{\eta}} (2 v_{\bar{\eta}}|Y_X|^2  + \sqrt{2} T_X)\;.
\end{eqnarray}
This matrix is diagonalized by $Z^R$
\begin{eqnarray}
Z^Rm^2_{\tilde{\nu}^R}Z^{R,\dag} = m_{2,\tilde{\nu}^R}^{dia}\;.
\end{eqnarray}

The mass matrix for CP-odd sneutrino $({\sigma}_{l}, {\sigma}_{r})$ is also deduced here
\begin{eqnarray}
m^2_{\tilde{\nu}^I} = \left(
\begin{array}{cc}
m_{{\sigma}_{l}{\sigma}_{l}} &m^T_{{\sigma}_{r}{\sigma}_{l}}\\
m_{{\sigma}_{l}{\sigma}_{r}} &m_{{\sigma}_{r}{\sigma}_{r}}\end{array}
\right)\;,
\end{eqnarray}
\begin{eqnarray}
&&m_{{\sigma}_{l}{\sigma}_{l}}= \frac{1}{8} \Big((g_{1}^{2} + g_{Y X}^{2} + g_{2}^{2}+  g_{Y X} g_{X})( v_{d}^{2}- v_{u}^{2})
+  2g_{Y X} g_{X}(v_{\eta}^{2}-v_{\bar{\eta}}^{2})\Big)
\nonumber\\&&\hspace{1.8cm}+\frac{1}{2} v_{u}^{2}{Y_{\nu}^{T}  Y_\nu}  + m_{\tilde{L}}^2\;,
 \\&&m_{{\sigma}_{l}{\sigma}_{r}} = \frac{1}{\sqrt{2} } v_uT_\nu -  v_u v_{\bar{\eta}} {Y_X  Y_\nu}
  - \frac{1}{2}v_d ({\lambda}_{H}v_S  + \sqrt{2} \mu )Y_\nu,\\&&
m_{{\sigma}_{r}{\sigma}_{r}}= \frac{1}{8} \Big((g_{Y X} g_{X}+g_{X}^{2})(v_{d}^{2}- v_{u}^{2})
+2g_{X}^{2}(v_{\eta}^{2}- v_{\bar{\eta}}^{2})\Big)- v_{\eta} v_S Y_X {\lambda}_{C}\nonumber \\&&\hspace{1.8cm}
+m_{\tilde{\nu}}^2 + \frac{1}{2} v_{u}^{2}|Y_\nu|^2+  v_{\bar{\eta}} (2 v_{\bar{\eta}}Y_X  Y_X  - \sqrt{2} T_X)\;.
\end{eqnarray}
This matrix is diagonalized by $Z^I$
\begin{eqnarray}
Z^Im^2_{\tilde{\nu}^I}Z^{I,\dag} = m_{2,\tilde{\nu}^I}^{dia}\;.
\end{eqnarray}

Mass matrix for charginos in the basis:($\tilde{W}^-$,$\tilde{H}_d^-$),($\tilde{W}^+$,$\tilde{H}_u^+$)
\begin{eqnarray}
m_{{\tilde{\chi}}^-}=
\left({\begin{array}{*{20}{c}}
M_2 & \frac{1}{\sqrt{2}}g_2v_u \\
\frac{1}{\sqrt{2}}g_2v_d & \frac{1}{\sqrt{2}}\lambda_Hv_S+\mu \\
\end{array}}
\right)\;,
\end{eqnarray}
The matrix is diagonalized by U and V
\begin{eqnarray}
U^*m_{{\tilde{\chi}}^-}V^\dag = m_{{\tilde{\chi}}^-}^{dia}.
\end{eqnarray}

The mass matrix for charged Higgs in the basis:($H_d^{-}$,$H_u^{+,*}$),($H_d^{-,*}$,$H_u^{+}$)
\begin{eqnarray}
m_{H^-}^2=
\left({\begin{array}{*{20}{c}}
m_{{H_d^{-}}H_d^{-,*}} & m_{H_u^{+,*}H_d^{-,*}}^{*} \\
m_{H_d^{-}H_u^{+}} & m_{H_u^{+,*}H_u^{+}} \\
\end{array}}
\right)\;,
\end{eqnarray}
\begin{eqnarray}
&&m_{{H_d^{-}}H_d^{-,*}}=\frac{1}{8}((g_2^2+g_X^2)v_d^2+(-g_X^2+g_2^2)v_u^2+(g_1^2+g_{YX}^2)(-v_u^2+v_d^2)-2g_X^2v_{\bar{\eta}}^2
\nonumber\\&&\hspace{1.8cm}+2(g_{YX}g_X(-v_{\bar{\eta}}^2-v_u^2+v_d^2+v_\eta^2)+g_X^2v_\eta^2)
\nonumber\\&&\hspace{1.8cm}+\frac{1}{2}(2\mid\mu\mid^2+2\sqrt{2}v_S\Re(\mu\lambda_H^*)+v_S^2\mid\lambda_H\mid^2\;,
\end{eqnarray}
\begin{eqnarray}
&&m_{H_d^{-}H_u^{+}}=\frac{1}{2}(2(\lambda_Hl_W^*+B_\mu)+\lambda_H(2\sqrt{2}v_SM_S^*-v_dv_u\lambda_H^*+v_\eta v_{\bar{\eta}}\lambda_C^*+\sqrt{2}v_ST_{\lambda_H}))
\nonumber\\&&\hspace{1.8cm}+\frac{1}{4}g_2^2v_dv_u\;,
\end{eqnarray}
\begin{eqnarray}
&&m_{H_u^{+,*}H_u^{+}}=\frac{1}{8}((-g_X^2+g_2^2)v_d^2+(g_2^2+g_X^2)v_u^2+(g_1^2+g_{YX}^2)(-v_d^2+v_u^2)-2g_X^2v_\eta^2
\nonumber\\&&\hspace{1.8cm}+2(g_{YX}g_X(-v_d^2-v_\eta^2+v_u^2+v_{\bar{\eta}}^2)+g_X^2v_{\bar{\eta}}^2))
\nonumber\\&&\hspace{1.8cm}+\frac{1}{2}(2\mid\mu\mid^2+2\sqrt{2}v_S\Re(\mu\lambda_H^*)+v_S^2\mid\lambda_H\mid^2)\;.
\end{eqnarray}
This matrix is diagonalized by $Z^+$
\begin{eqnarray}
Z^+m_{H^-}^2Z^{+,\dag} = m_{2,H^-}^{dia}\;.
\end{eqnarray}

The mass matrix for neutralino in the basis($\lambda_{\tilde{B}}$,$\tilde{W}^0$,$\tilde{H}_d^0$,$\tilde{H}_u^0$,$\lambda_{\tilde{X}}$,$\tilde{\eta}$,$\tilde{\bar{\eta}}$,$\tilde{s}$) is
\begin{eqnarray}
m_{\tilde{\chi}^0}=
\left({\begin{array}{*{20}{c}}
M_1 & 0 & -\frac{g_1}{2}v_d & \frac{g_1}{2}v_u & M_{{BB}^{\prime}} & 0 & 0 & 0 \\
0 & M_2 & \frac{g_2}{2}v_d & -\frac{g_2}{2}v_u & 0 & 0 & 0 & 0 \\
-\frac{g_1}{2}v_d & \frac{g_2}{2}v_d & 0 & m_{{\tilde{H}_u^0}{\tilde{H}_d^0}} & m_{\lambda_{\bar{X}}\tilde{H}_d^0} & 0 & 0 & -\frac{\lambda_Hv_u}{\sqrt{2}} \\
\frac{g_1}{2}v_u & -\frac{g_2}{2}v_u & m_{{\tilde{H}_d^0}{\tilde{H}_u^0}} & 0 & m_{\lambda_{\bar{X}}{\tilde{H}_u^0}} & 0 & 0 & -\frac{\lambda_Hv_d}{\sqrt{2}} \\
M_{{BB}^\prime} & 0 & m_{\tilde{H}_d^0\lambda_{\bar{X}}} & m_{\tilde{H}_u^0\lambda_{\bar{X}}} & M_{BL} & -g_X{v_\eta} & g_Xv_{\bar{\eta}} & 0 \\
0 & 0 & 0 & 0 & -g_X{v_\eta} & 0 & \frac{1}{\sqrt{2}}\lambda_Cv_S & \frac{1}{\sqrt{2}}\lambda_Cv_{\bar{\eta}} \\
0 & 0 & 0 & 0 & g_Xv_{\bar{\eta}} & \frac{1}{\sqrt{2}}\lambda_Cv_S & 0 & \frac{1}{\sqrt{2}}\lambda_Cv_\eta \\
0 & 0 & -\frac{1}{\sqrt{2}}\lambda_Hv_u & -\frac{1}{\sqrt{2}}\lambda_Hv_d & 0 & \frac{1}{\sqrt{2}}\lambda_Cv_{\bar{\eta}} &  \frac{1}{\sqrt{2}}\lambda_Cv_\eta & m_{\tilde{s}\tilde{s}} \\
\end{array}}
\right)\;,
\end{eqnarray}
\begin{eqnarray}
&&m_{{\tilde{H}_d^0}{\tilde{H}_u^0}}=-\frac{1}{\sqrt{2}}\lambda_Hv_S - \mu, ~~ m_{{\tilde{H}_d^0}\lambda_{\bar{X}}}=-\frac{1}{2}(g_{YX}+g_X){v_d},
\nonumber\\&&\ m_{\tilde{H}_u^0\lambda_{\bar{X}}}=\frac{1}{2}(g_{YX}+g_X)v_u, ~~ m_{\tilde{s}\tilde{s}}=2M_S+\sqrt{2}\kappa v_S\;.
\end{eqnarray}
This matrix is diagonalized by $N$,
\begin{eqnarray}
N^*m_{{\tilde{\chi}}^0}N^\dag=m_{{\tilde{\chi}}^0}^{dia}\;.
\end{eqnarray}

\end{document}